\documentclass{scrartcl}
\usepackage{amsmath,amssymb}
\usepackage{amsthm}
\usepackage{slashed}
\usepackage[dvipdfmx]{graphicx}
\usepackage[headsep=15pt,head=80pt,foot=20pt,margin=60pt]{geometry}

\newcommand{\Group}[2]{{ \hbox{{\itshape{#1}}($#2$)} }}
\newcommand{\U}[1]{\Group{U\kern0.05em}{#1}}
\newcommand{\SU}[1]{\Group{SU\kern0.1em}{#1}}
\newcommand{\SL}[1]{\Group{SL\kern0.05em}{#1}}
\newcommand{\Sp}[1]{\Group{Sp\kern0.05em}{#1}}
\newcommand{\SO}[1]{\Group{SO\kern0.1em}{#1}}

\newcommand{\mybar}[1]%
    {{\kern 0.8pt\overline{\kern -0.8pt#1\kern -0.8pt}\kern 0.8pt}}
\newcommand{\sla}[1]%
    {{\raise.15ex\hbox{$/$}\kern-.57em #1}}
\newcommand{\roughly}[1]%
    {{ \mathrel{\raise.3ex\hbox{ $#1$\kern-.75em\lower1ex\hbox{$\sim$}} } }}

\newcommand{\nop}[1]{:\kern-.3em#1\kern-.3em:}

\newcommand{\del}{\partial}

\newcommand{\covfb}{\overleftrightarrow{D}}
\newcommand{\al}{\ensuremath{\alpha}}
\newcommand{\be}{\ensuremath{\beta}}

\newcommand{\de}{\ensuremath{\delta}}

\newcommand{\la}{\ensuremath{\lambda}}

\newcommand{\rh}{\ensuremath{\rho}}
\newcommand{\si}{\ensuremath{\sigma}}

\newcommand{\ph}{\ensuremath{\phi}}

\newcommand{\GeV}{\ensuremath{\mathrm{~GeV}}%
}
\newcommand{\TeV}{\ensuremath{\mathrm{~TeV}}%
}

\newcommand{\hc}{\mathrm{H.c.}} 
\newcommand{\n}{\notag \\}
\newcommand{\mcl}[1]{\mathcal{#1}}

\numberwithin{equation}{section}
\numberwithin{figure}{section}
\numberwithin{table}{section}

\begin{document}
\begin{titlepage}

\begin{flushright}
UG-FT-311/14\\
CAFPE-181/14
\end{flushright}
\vskip 2cm

\begin{center}
{\Large \bfseries
 Derivative Interactions and Perturbative UV Contributions\\in $N$ Higgs Doublet Models
}

\vskip 1.2cm

Yohei Kikuta$^\sharp$, \hspace{1em}
Yasuhiro Yamamoto$^\flat$

\bigskip
\bigskip

\begin{center}
$^\sharp$ 
  \textit{KEK Theory Center, KEK,}\\
  \textit{and}\\
  \textit{Department of Particle and Nuclear Physics, The Graduate University for Advanced Studies,}\\
  \textit{Tsukuba, JAPAN}\\[1em]
$^\flat$
  \textit{Deportamento de Fisica Teorica y del Cosmos, Facultad de Ciencias and CAFPE,}\\
  \textit{Universidad de Granada, Granada, SPAIN}\\
\end{center}

\vskip 1.5cm

\textbf{abstract}
\end{center}

\medskip
\noindent

We study the Higgs derivative interactions on models including arbitrary number of the Higgs doublets.
These interactions are generated by two ways.
One is higher order corrections of composite Higgs models, and the other is integration of heavy scalars and vectors.
In the latter case, three point couplings between the Higgs doublets and these heavy states are the sources of the derivative interactions.
Their representations are constrained to couple with the doublets.
We explicitly calculate all derivative interactions generated by integrating out.
Their degrees of freedom and conditions to impose the custodial symmetry are discussed.
We also study the vector boson scattering processes with a couple of two Higgs doublet models to see experimental signals of the derivative interactions. 
They are differently affected by each heavy field.

\bigskip
\vfill
\end{titlepage}

\tableofcontents
\section{Introduction}
\label{SecIntro}
The Standard Model (SM) has been completed by the discovery of the Higgs boson~\cite{RefHiggs}.
However, the SM does not explain the mechanism of the electroweak symmetry breaking (EWSB).
The symmetry is broken by hand with the negative mass square of the Higgs doublet.
In order to find some clues of the origin of the EWSB, physics of the Higgs sector has been investigated well.

Physics of the higher energy region can be described with higher dimensional operators at the lower energy region.
Their effects appear as smoking gun signals of new physics.
For the Higgs sector, the Higgs derivative interactions, which consist of two derivatives and four Higgs doublets, have attracted interest as operators which shed light on a new structure behind the Higgs sector.
These operators modify the normalization of the Higgs boson, so that they ubiquitously affect the Higgs physics.
In particular, the cross sections of the vector boson scatterings are enhanced by this effect because they violate the unitarity condition of these processes.

They have been studied well by Ref.~\cite{RefSilh} in the context of composite Higgs models.
In these models, the Higgs doublet is embedded in nonlinear sigma models.
The derivative interactions are given by the expansion of the kinetic term. 
Their effects in future colliders has been studied by many papers, e.g. Refs.~\cite{RefSdhp,RefIlc}.

The derivative interactions are also generated by integrating out heavy scalars and vectors.
Three point interactions between the Higgs bilinears and heavy bosons are the sources of these effective interactions.
They also appear in composite Higgs models.
Since the representations of the heavy fields are constrained to couple with the bilinears, the interactions generated by integrating out have qualitatively different features from composite ones~\cite{RefLrv,RefCmpr}.

Introducing additional Higgs doublets is a popular way to extend the Higgs sector.
In particular, the minimal extension, the two Higgs doublet models, are studied well.
The derivative interactions also appear in these models because of other heavy bosons and the composite nature.
For example, an explicit model is studied by Ref.~\cite{RefN2hdm}.
Some general features of the Higgs derivative interactions have been investigated by Refs.~\cite{RefKoy,RefKy} in the context of composite $N$ Higgs doublet models.

In this paper, we calculate the derivative interactions generated by the heavy boson integrations for models including an arbitrary number of the Higgs doublets.
The given derivative interactions are constrained because we assume models have perturbative UV completions\footnotemark.
\footnotetext{
 Here a perturbative UV completion does not mean a renormalizable theory, which is valid up to very high scale like the GUT scale or the Planck scale.
 We consider possible models just behind the electroweak scale including TeV scalar and vectors.
 For instance, composite vector resonances are also considered as heavy vectors.
}
We investigate some features appearing in the vector boson scatterings on the two Higgs doublet models.

Recently, some experimental results from ATLAS and CMS implied new bosons around 2\TeV~\cite{Ref2tev}.
Their couplings with the Higgs boson could help us to investigate the origin of the EWSB.
The low energy structure of couplings between these new particles and the doublets are explicitly shown in this manuscript.

This paper is organized as follows.
After this introduction, we show the derivative interactions generated by the heavy boson integrations in Sect.~\ref{SecInt}.
We also discuss conditions imposing the custodial symmetry to the interactions and terms contributing the oblique parameters there.
In Sect.~\ref{SecProp}, we investigate some properties of the given derivative interactions.
For the case of the $N$ Higgs doublet models, we compare the degrees of freedom (DOFs) of the derivative interactions we have obtained and the general effective Lagrangian of the derivative interactions which respect only the SM symmetry.
For the two Higgs doublet models, we discuss how the contribution occurs as regards the vector boson scattering processes.
We present our conclusions in Sect.~\ref{SecCon}.
The potential terms simultaneously generated by the heavy scalar integrations and  the conditions to impose the custodial symmetry on the derivative interactions are shown in Appendices~\ref{AppPot} and \ref{AppCust}, respectively.
\section{Integrating out and effective Lagrangian}
\label{SecInt}
We explicitly calculate the Higgs derivative interactions given by the integration of heavy scalar bosons and vector bosons.
The derivative interactions are generated by three point interactions among them.
First, we explain the procedure and the notation used in this paper.
Afterwards, the explicit forms of the derivative interactions are shown with a conventional operator base.
It is a straightforward generalization of Ref.~\cite{RefLrv}.

The contributions of the heavy scalars appear in various sets of the Higgs flavor indices.
On the other hand, those of the heavy vectors are strongly constrained because we assume they interact with doublets through kinetic terms, namely, flavor diagonal terms.

We also discuss conditions to preserve the custodial symmetry in the derivative interactions based on Ref.~\cite{RefKy}.
These conditions are discussed in terms of the Wilson coefficients, namely, the tree level, because 1-loop contributions can also be produced by other sectors. 
The 1-loop calculation has been studied in Ref.~\cite{Ref1loop}

The disappearance of the Wilson coefficients obtained in this section is related to the unitarity sum rule.
A consequence of the sum rule lead to the custodial symmetry and other relations as studied in Refs.~\cite{RefUnitaritysum}.
This rigidity can also be observed below.

\subsection{Integration of heavy bosons}
First, we study the Higgs derivative interactions generated by the heavy scalars.
Since we focus on the three point interactions, these heavy scalars can be classified into four kinds of representations with respect to SU(2)$_L\times$U(1)$_Y$.
They are the SU(2)$_L$ singlet or triplet with the hypercharge of 0 or 1.
Hereafter, the singlet and the triplet are, respectively, referred to as \textbf{1} and \textbf{3}, and their hypercharges are shown as their subscripts.
Even if there are many fields for a representation, their differences are merged in terms of an effective interaction, except for some special situations mentioned later in this section.
We calculate the effective interactions using a scalar field for each representation.

Our discussion is based on the following Lagrangian:
\begin{align}
 \mcl{L}^S =&
  -\frac{m_0^2}{2} \ph_0^2 
  +m_0 \ph_0 \left( 
    \sum_i \la_0^{ii} I^0_{ii}
   +\sum_{i<j} (\la_0^{ij} I^0_{ij} +\hc) 
  \right) \n &
  -m_S^2 \ph_S^\dag \ph_S
  +m_S \sum_{i<j} \left( \la_S^{ij} \ph_S^\dag I^S_{ij} +\hc \right) \n &
  -\frac{m_L^2}{2} \ph_L^a \ph_L^a
  +m_L \ph_L^a \left(
    \sum_i \la_L^{ii} I^{La}_{ii} 
   +\sum_{i<j} (\la_L^{ij} I^{La}_{ij} +\hc ) 
	\right) \n &
  -m_T^2 \ph_T^{a\dag} \ph_T^a 
  +m_T \biggl(\ph_T^{a\dag} \left(
    \sum_i \frac{\la_T^{ii}}{2} I^{Ta}_{ii}
   +\sum_{i<j} \la_T^{ij} I^{Ta}_{ij}
	\right) +\hc  \biggr),
\end{align}
where
\begin{align}
 I^0_{ij} =& H_i^\dag H_j, \\
 I^S_{ij} =& H_i^T \si^2 H_j, \\
 I^{La}_{ij} =& H_i^\dag \si^a H_j, \\
 I^{Ta}_{ij} =& H_i^T \si^2 \si^a H_j,
\end{align}
and $\ph_0$, $\ph_S$, $\ph_L^a$, and $\ph_T^a$ are the heavy scalar fields of the $\textbf{1}_0$, $\textbf{1}_1$, $\textbf{3}_0$, and $\textbf{3}_1$ representations, respectively\footnotemark.
\footnotetext{
 The triplet scalar fields, $\textbf{3}_{0,1}$, have vacuum expectation values which explicitly violate the custodial symmetry.
 If we write them as $v_\ph$, they roughly satisfy the relation
 \begin{align}
  \la \frac{v}{M} \sim \frac{v_\ph}{v},
 \end{align}
 where $\la$, $v$, and $M$ stand for the three point coupling, the vacuum expectation value of the doublet, and the mass of the triplet scalar, respectively.
 Because of the electroweak precision measurement, $v_\ph /v \sim 0.01$ for each triplet.
 This means $M$ is a few tens \TeV for $\la =1$, so that their effects are much smaller than the expected limits written in Ref.~\cite{RefIlc}.
 However, the vacuum expectation values of the $\textbf{3}_0$ scalar and the $\textbf{3}_1$ scalar can be canceled each other, i.e., the alignment limit of the Georgi--Machacek model.
 In this case, $v_\ph /v$ can be a few tens \GeV, so $M$ can be about 1 \TeV.
 These discussions are implicitly assumed in Ref.~\cite{RefLrv}.
 The current situation of the latter case has been studied by Ref.~\cite{RefTriplet}.
}
Using the symmetry of the indices for the doublets in the above operators, the couplings satisfy the following relations,
\begin{align}
 \la_0^{ij} =& \la_0^{ji\ast}, \\
 \la_S^{ij} =& -\la_S^{ji}, \\
 \la_L^{ij} =& \la_L^{ji\ast}, \\
 \la_T^{ij} =& \la_T^{ji}.
\end{align}
The flavor diagonal components of the $\textbf{1}_0$ and $\textbf{3}_0$ couplings, $\la_0^{ii}$ and $\la_L^{ii}$, are real.
The couplings of the $\textbf{1}_1$ scalars, $\la_S^{ij}$, are antisymmetric under the exchange of indices.
Hence, this coupling disappears in the one Higgs doublet models.

Integrating out the heavy scalars roughly generates the effective interactions as follows:
\begin{align}
 -\frac{m_L^2}{2} \ph_L^a \ph_L^a 
 +\la_L m_L \ph_L^a H^\dag \si^a H 
 \Rightarrow &
 \frac{\la_L^2 m_L^2}{2} (H^\dag \si^a H) 
  \frac{1}{\del^2 +m_L^2} (H^\dag \si^a H) \\
 = &
 \frac{\la_L^2}{2} (H^\dag \si^a H)
   \left( 1-\frac{\del^2}{m_L^2} +\cdots \right) (H^\dag \si^a H).
\end{align}
In the large parenthesis of the last line, the first term contributes to the potential, and the second term gives us the derivative interactions we are interested in.
The generated potential terms are shown in Appendix~\ref{AppPot}.

Following the above procedure, the given derivative interactions are written as
\begin{align}
 \mcl{L}_\text{eff}^S =&
   \frac{1}{2m_0^2} \left( \del \left(
	  \sum_i \la_0^{ii} I^0_{ii}
    +\sum_{i<j} (\la_0^{ij} I^0_{ij} +\hc) 
   \right) \right)^2 \n &
  +\frac{1}{m_S^2} 
    \sum_{i<j,k<l}
     \del \left( \la_S^{ij} I^S_{ij} +\hc \right) 
     \del \left( \la_S^{kl} I^{S\dag}_{kl} +\hc \right) \n &
  +\frac{1}{2m_L^2} \left( \del \left(
     \sum_i \la_L^{ii} I^{La}_{ii} 
    +\sum_{i<j} (\la_L^{ij} I^{La}_{ij} +\hc )
   \right) \right)^2 \n &
  +\frac{1}{m_T^2} 
	  \del \left(
       \sum_i \frac{\la_T^{ii}}{2} I^{Ta}_{ii} +\sum_{i<j} \la_T^{ij} I^{Ta}_{ij}
	  \right)
	  \del \left( 
       \sum_k \frac{\la_T^{kk\ast}}{2} I^{Ta\dag}_{kk} +\sum_{k<l} \la_T^{kl\ast} I^{Ta\dag}_{kl}
	  \right).
\end{align}
These interactions can be expressed with the following four kinds of operators:
\begin{align}
 & H_i^\dag H_j (\del H_j)^\dag (\del H_k), \\
 & \del (H_i^\dag H_j) (H_k^\dag \covfb H_l), \\
 & \del (H_i^\dag H_j) \del (H_k^\dag H_l), \\
 & (H_i^\dag \covfb H_j ) (H_k^\dag \covfb H_l),
\end{align}
where $H^\dag_i \covfb H_j$ stands for $H^\dag_i (D H_j) -(D H^\dag_i ) H_j$ and the Lorentz indices are suppressed.
As is well known, the first two kinds of operators can be eliminated with the field redefinition\footnotemark.
\footnotetext{
 This procedure simultaneously introduces higher dimensional operators in the Yukawa couplings.
 They are not specified in this paper.
}
We use the rest of the operators to express the given derivative interactions with the notation below,
\begin{align}
 O^H_{ijkl} =& 
   \frac{1}{1 +\de_{ik} \de_{jl}} \del (H_i^\dag H_j) \del (H_k^\dag H_l) \\
 O^T_{ijkl} =& 
   \frac{1}{1 +\de_{ik} \de_{jl}} (H_i^\dag \covfb H_j)(H_k^\dag \covfb H_l).
\end{align}
It is convenient to classify these operators for the combinations of the indices as follows:
\begin{description}
 \item[type I]: all doublets have the same flavor, e.g. $\del (H_i^\dag H_i) \del (H_i^\dag H_i)$;
 \item[type II]: one of doublets is different from the others, e.g. $\del (H_i^\dag H_i) \del (H_i^\dag H_j)$;
 \item[type III]: two flavors of doublets are included by two each e.g. $\del (H_i^\dag H_i) \del (H_j^\dag H_j)$;
 \item[type IV]: three flavors of doublets are included in operator e.g. $\del (H_i^\dag H_i) \del (H_j^\dag H_k)$;
 \item[type V]: all doublets have different indices e.g. $\del (H_i^\dag H_j) \del (H_k^\dag H_l)$.
\end{description}

The integration of the heavy vector fields are similar to that of the heavy scalar fields.
We study the following Lagrangian to derive the Higgs derivative interactions\footnotemark:
\footnotetext{
 Fermion currents can also couple with the heavy vector fields.
 However, they can be eliminated by the SM equation of motion as mentioned in Ref.~\cite{RefLrv}.
}
\begin{align}
 \mcl{L}^V =&
   \frac{m_0^2}{2} V_0\mu V_{0\mu}
  +V_{0\nu} \left(
    \sum_i g_0^{ii} J^0_{ii\nu}
   +g' (\rh_0 \del_\mu B^{\mu\nu})
   \right) \n &
  +m_S^2 V_S^{\mu\dag} V_{S\mu}
  +\biggl( V_S^{\mu\dag} \left(
    \sum_i \frac{g_S^{ii}}{2} J^S_{ii\mu} 
   \right) +\hc \biggr) \n &
  +\frac{m_L^2}{2} V_L^{a\mu} V_{L\mu}^a 
  + V_{L\nu}^a  \biggl(
    \sum_i g_L^{ii} J^{La}_{ii\nu}
   +g \rh_L (D_\mu W^{\mu\nu})^a \biggr),
\end{align}
where the Higgs currents are defined as
\begin{align}
 J^0_{ii} =& i H_i^\dag \covfb H_i ,\\
 J^S_{ii} =& i H_i^T \si^2 \covfb H_i ,\\
 J^{La}_{ii} =& i H_i^\dag \si^a \covfb H_i .
\end{align}
The couplings, $g_{0,L}^{ii}$ are real and $g_S^{ii}$ can be complex.
Since we assume the couplings with the vector fields appear from the kinetic terms, the vector of the $\textbf{3}_1$ representation, which only interacts with the flavor off diagonal currents, is not included.
The $\textbf{1}_1$ vector can be introduced as the $\textbf{1}_0$ and the $\textbf{3}_0$ in terms of the O(4) representation of the Higgs boson.
It is embedded in a part of the SU(2)$_R$.
Further details are described in Ref.~\cite{RefLrv}.

Integrating out the heavy vectors, we obtain the effective Lagrangian which consists of the squared currents,
\begin{align}
 \mcl{L}_\text{eff}^V =&
  -\frac{1}{2m_0^2} \biggl(
	  \sum_i g_0^{ii} J^0_{ii}
	 +g' \rh_0 \del \cdot B 
  \biggr)^2 \n &
  -\frac{1}{m_S^2} 
	  \left( 
      \sum_i \frac{g_S^{ii}}{2} J^S_{ii}
	  \right) \cdot
	  \left(
      \sum_j \frac{g_S^{jj\ast}}{2} J^{S\dag}_{jj}
	  \right) \n &
   -\frac{1}{2m_L^2} \biggl(
	  \sum_i g_L^{ii} J^{La}_{ii}
    +g \rh_L (\del \cdot W)^a
   \biggr)^2 .
\label{EqVec}
\end{align}
With the effective operators of $O^H$ and $O^T$, only the type I and the type III effective operators appear due to their flavor diagonal couplings.
In the case of the heavy vectors, the derivative interactions are the leading contributions of integrating out.
They do not give us additional potential terms.

The above effective Lagrangian also produces higher dimensional operators including the gauge currents, $\del \cdot B$ and $(D \cdot W)^a$.
They contribute to the oblique parameters introduced by Refs.~\cite{RefPT,RefEWPM}.
We will take care of these tree level effects since they are usually too large to compensate with contributions of the other sectors because of the same reason as with the custodial symmetry.
\subsection{Type I}
Integrating out the heavy scalars, the following type I derivative interactions are obtained:
\begin{align}
 \mcl{L}_\text{I}^S =&
  \frac{\la_0^{ii 2}}{2m_0^2}\left( \del I^0_{ii} \right)^2
 +\frac{\la_L^{ii 2}}{2m_L^2}\left( \del I^{La}_{ii} \right)^2
 +\frac{\la_T^{ii}\la_T^{ii\ast}}{4m_T^2} \left( \del I^{Ta}_{ii} \right) \left( \del I^{Ta\ast}_{ii} \right) \\
\Rightarrow &
  \left(
   \frac{\la_0^{ii 2}}{m_0^2}
  -\frac{2\la_L^{ii 2}}{m_L^2}
  -\frac{\la_T^{ii} \la_T^{ii\ast}}{2m_T^2}
 \right) O^H_{iiii}
 +\left(
   \frac{\la_L^{ii2}}{m_L^2} -\frac{\la_T^{ii} \la_T^{ii\ast}}{2m_T^2}
 \right) O^T_{iiii} .
\end{align}

Since the operator $O^T_{iiii}$ violates the custodial symmetry, the following condition is required:
\begin{align}
 \frac{\la_L^{ii2}}{m_L^2} =\frac{\la_T^{ii}\la_T^{ii\ast}}{2m_T^2}.
\end{align}
The $\textbf{1}_0$ scalar preserves the symmetry.
It also happens for the other types of the derivative interactions as shown in the rest of this section.

If we suppose that $\la_T^{ii}$ is real\footnotemark, the condition can be written,
\begin{align}
  \frac{\la_L^{ii}}{m_L} = \pm \frac{\la_T^{ii}}{\sqrt{2}m_T}.
	\label{EqCustI}
\end{align}
Their signs can be independently chosen for each doublet.
This relation means that different representations of SU(2)$_R$ collectively recover the custodial SO(4) symmetry.
This situation is expected if they are embedded in a larger multiplet respecting the symmetry in UV completions.
\footnotetext{
 This assumption to show a simple situation may looks artificial.
 However, imaginary parts of couplings should satisfy too many relations to preserve the custodial symmetry as shown below.
 Even if they can survive the relations, couplings with the Standard Model fermions not studied in this paper give strong constraints on these CP violating phases; see e.g. Ref.~\cite{RefAbe}.
}

Likewise, in the above calculation, the heavy vectors produce the derivative interactions,
\begin{align}
 \mcl{L}^V_\text{I} =&
  -\frac{g_0^{ii2}}{2m_0^2} J^0_{ii}{}^2
	-\frac{g_S^{ii}g_S^{ii\ast}}{4m_S^2}J^S_{ii} J^{S\dag}_{ii}
	-\frac{g_L^{ii2}}{2m_L^2}J^{La}_{ii}{}^2 \\
\Rightarrow &
  3\left(
    \frac{g_S^{ii} g_S^{ii\ast}}{2m_S^2}
   +\frac{g_L^{ii 2}}{m_L^2}
  \right) O^H_{iiii}
 +\left(
    \frac{g_0^{ii 2}}{m_0^2}
	-\frac{g_S^{ii} g_S^{ii\ast}}{2m_S^2}
  \right) O^T_{iiii} .
\end{align}

The custodial symmetry is recovered by eliminating the coefficient of $O^T$ with the condition,
\begin{align}
  \frac{g_0^{ii2}}{m_0^2} = \frac{g_S^{ii}g_S^{ii\ast}}{2m_S^2} .
\end{align}
For the case of the scalars, the condition is imposed between the \textbf{3} representations.
However, in this case, the \textbf{1} representations are related each other.

As discussed in the heavy scalar case, if $g_S^{ii}$ is real, the sizes of the interactions are fixed up to signs,
\begin{align}
 \frac{g_0^{ii}}{m_0} = \pm \frac{g_S^{ii}}{\sqrt{2}m_S}.
 \label{EqCustVI}
\end{align}
The $\textbf{3}_0$ vector does not violate the custodial symmetry such as the $\textbf{1}_0$ scalar.
The results obtained here are the same as that given by Ref.~\cite{RefLrv}.
\subsection{Type II}
The effective Lagrangian of the type II derivative interactions are similar to that of the type I,
\begin{align}
 \mcl{L}_\text{II}^S =&
  \frac{\la_0^{ii}\la_0^{ij}}{m_0^2} \left( \del I^0_{ii} \right) \left( \del I^0_{ij} \right)
 +\frac{\la_L^{ii}\la_L^{ij}}{m_L^2} \left( \del I^{La}_{ii} \right) \left( \del I^{La}_{ij} \right)
 +\frac{\la_T^{ii}\la_T^{ij\ast}}{2m_T^2} \left( \del I^{Ta}_{ii} \right) \left( \del I^{Ta\ast}_{ij}  \right) +\hc \\
=&
  \left(
   \frac{\la_0^{ii} \la_0^{ij}}{m_0^2} -\frac{2\la_L^{ii} \la_L^{ij}}{m_L^2}
  -\frac{\la_T^{ij} \la_T^{ii\ast}}{2m_T^2}
  \right) O^H_{iiij}
 +\left(
   \frac{\la_L^{ii} \la_L^{ij}}{m_L^2} -\frac{\la_T^{ij} \la_T^{ii\ast}}{2m_T^2}
  \right) O^T_{iiij}
 +\hc .
\end{align}
As discussed in Ref.~\cite{RefKoy}, the above Wilson coefficients have to be real to preserve the custodial symmetry.
Besides, the following condition is required because the coefficient of $O^T_{iiij}$ violates the symmetry:
\begin{align}
   \frac{\la_L^{ii} \la_L^{ij}}{m_L^2} =\frac{\la_T^{ij} \la_T^{ii\ast}}{2m_T^2}.
\end{align}
Assuming the three point couplings are real, with Eq.~\eqref{EqCustI}, the above relation becomes
\begin{align}
  \frac{\la_L^{ij}}{m_L} = \pm \frac{\la_T^{ij}}{\sqrt{2}m_T}.
  \label{EqCustII}
\end{align}
The relative sign is the same as in Eq.~\eqref{EqCustI} for any combination of the indices.
\subsection{Type III}
\label{SecTypeIII}
The type III derivative operators given by the heavy scalars are shown now,
\begin{align}
 \mcl{L}^S_\text{III} =&
   \frac{1}{2m_0^2} \left(\la_0^{ij}\del I^0_{ij} +\hc\right)^2
	+\frac{\la_0^{ii}\la_0^{jj}}{m_0^2} \left(\del I^0_{ii}\right)\left(\del I^0_{jj}\right)
	+\frac{\la_S^{ij}\la_S^{ij\ast}}{m_S^2} \left(\del I^S_{ij}\right)\left(\del I^{S\dag}_{ij}\right) \n &
	+\frac{1}{2m_L^2}\left(\la_L^{ij}\del I^{La}_{ij} +\hc\right)^2
	+\frac{\la_L^{ii}\la_L^{jj}}{m_L^2}\left(\del I^{La}_{ii}\right) \left(\del I^{La}_{jj}\right) \n &
	+\frac{\la_T^{ij}\la_T^{ij\ast}}{m_T^2} \left(\del I^{Ta}_{ij}\right)\left(\del I^{Ta\dag}_{jj} \right)
	+\left( \frac{\la_T^{ii}\la_T^{jj\ast}}{4m_T^2} \left(\del I^{Ta}_{ii}\right)\left(\del I^{Ta\dag}_{jj}\right) +\hc \right) \\
\Rightarrow &
  \left(
    \frac{\la_0^{ii} \la_0^{jj}}{m_0^2}
	-\frac{\la_S^{ij} \la_S^{ij\ast}}{2m_S^2}
   -\frac{\la_L^{ij} \la_L^{ij\ast} +\la_L^{ii} \la_L^{jj}}{m_L^2}
   -\frac{\la_T^{ij} \la_T^{ij\ast}}{2m_T^2}
  \right) O^H_{iijj} \n &
 +\left(
    \frac{\la_0^{ij} \la_0^{ij\ast}}{m_0^2}
   +\frac{\la_S^{ij} \la_S^{ij\ast}}{2m_S^2}
   -\frac{\la_L^{ij} \la_L^{ij\ast} +\la_L^{ii} \la_L^{jj}}{m_L^2}
   -\frac{\la_T^{ij} \la_T^{ij\ast}}{2m_T^2}
  \right) O^H_{ijji} \n &
 +\Bigl( \left(
    \frac{\la_0^{ij 2}}{m_0^2}
   -\frac{2\la_L^{ij 2}}{m_L^2}
   -\frac{\la_T^{jj} \la_T^{ii\ast}}{2m_T^2}
  \right) O^H_{ijij} +\hc \Bigr) \n &
 +\left(
   -\frac{\la_S^{ij} \la_S^{ij\ast}}{2m_S^2}
   +\frac{\la_L^{ij} \la_L^{ij\ast}}{m_L^2}
   -\frac{\la_T^{ij} \la_T^{ij\ast}}{2m_T^2}
  \right) O^T_{iijj} \n &
 +\left(
    \frac{\la_S^{ij} \la_S^{ij\ast}}{2m_S^2}
   +\frac{\la_L^{ii} \la_L^{jj}}{m_L^2}
   -\frac{\la_T^{ij} \la_T^{ij\ast}}{2m_T^2}
  \right) O^T_{ijji}
 +\Bigl( \left(
    \frac{\la_L^{ij 2}}{m_L^2}
	-\frac{\la_T^{jj} \la_T^{ii\ast}}{2m_T^2}
  \right) O^T_{ijij} +\hc \Bigr).
\end{align}
The couplings with the $\textbf{1}_1$ scalars appear of this type.

To impose the custodial symmetry, the coefficients of $O^H_{ijij}$ and $O^T_{ijij}$ have to be real.
We also need the following two nontrivial relations to preserve the symmetry:
\begin{align}
 \frac{\la_T^{ii} \la_T^{jj}}{2m_T^2} =&
   \frac{\la_L^{ii} \la_L^{jj}}{m_L^2} +\frac{\la_S^{ij2}}{3m_S^2}, \\
 \frac{\la_T^{ij2}}{2m_T^2} =&
   \frac{\la_L^{ij2}}{m_L^2} -\frac{\la_S^{ij2}}{6m_S^2},
\end{align}
where, for simplicity, it is supposed that all couplings are real.
The conditions without this simplification are shown in Appendix~\ref{AppCust}.
The couplings with the $\textbf{1}_1$ scalars are also required to preserve the symmetry, unlike the type I and the type II operators.

The type I, II, and III interactions appear for any pair of indices in models including more than two doublets.
Using Eqs.~\eqref{EqCustI} and \eqref{EqCustII}, it is found that the $\textbf{1}_1$ couplings, $\la_S^{ij}$, disappear and the relative signs between $\la_L$ and $\la_T$ are the same for any pair of the indices.
If the heavy scalars have only flavor off diagonal couplings, the relative signs are still free.
Another way not to fix the relative signs is as follows.
We divide the doublets into several groups, and each group couples with different heavy scalars for each representation.
In this case, the relative signs can be changed for the different groups.

The following type III derivative operators are obtained by integrating out the heavy vectors:
\begin{align}
 \mcl{L}_\text{III}^V =&
	-\frac{g_0^{ii}g_0^{jj}}{m_0^2}J^0_{ii} J^0_{jj}
	-\frac{1}{4m_S^2} \left( g_S^{ii}g_S^{jj\ast} J^S_{ii}J^{S\dag}_{jj} +\hc \right)
	-\frac{g_L^{ii}g_L^{jj}}{m_L^2} J^{La}_{ii} J^{La}_{jj} \\
\Rightarrow &
  3\left(
   \frac{g_L^{ii} g_L^{jj}}{m_L^2}
  \right) O^H_{ijji}
 +3\biggl( \left(
   \frac{g_S^{jj} g_S^{ii\ast}}{2m_S^2}
  \right) O^H_{ijij} +\hc \biggr) \n &
 +\left(
   \frac{g_0^{ii} g_0^{jj}}{m_0^2}
  -\frac{g_L^{ii} g_L^{jj}}{m_L^2}
  \right) O^T_{iijj}
 +\left(
   \frac{g_L^{ii} g_L^{jj}}{m_L^2}
  \right) O^T_{ijji}
 -\biggl( \left(
   \frac{g_S^{jj} g_S^{ii\ast}}{2m_S^2}
  \right) O^T_{ijij} +\hc \biggr).
\end{align}
Assuming that the vector couplings are real, the two conditions of the custodial symmetry for the Wilson coefficients become the same.
The condition is similar to those of the type I and II,
\begin{align}
 \frac{g_0^{ii} g_0^{jj}}{m_0^2} =&
 \frac{g_S^{ii} g_S^{jj}}{2m_S^2}.
\end{align}
Using this condition, the relative signs written in Eq.~\eqref{EqCustVI} become the same for the different Higgs doublets.
If the heavy vectors couple with only two of the doublets, the imaginary parts of $g_S$ can be kept after imposing the custodial symmetry.
\subsection{Type IV}
The type IV derivative operators are calculated as follows:
\begin{align}
 \mcl{L}_\text{IV}^S =&
   \frac{1}{m_0^2} \left(\la_0^{ij}\del I^0_{ij} +\hc \right) \left(\la_0^{ik}\del I^0_{ik} +\hc \right) 
	+\frac{\la_0^{ii}}{m_0^2}\left(\del I^0_{ii}\right) \left(\la_0^{jk} \del I_{jk}^0 +\hc\right) \n &
	+\left(\frac{\la_S^{ij}\la_S^{ik\ast}}{m_S^2} (\del I^S_{ij})(\del I^{S\ast}_{ik}) +\hc \right) \n &
  +\frac{1}{m_L^2} \left(\la_L^{ij}\del I^{La}_{ij} +\hc\right) \left(\la_L^{ik}\del I^{La}_{ik} +\hc\right) 
	+\frac{\la_L^{ii}}{m_L^2}\left(\del I^{La}_{ii}\right) \left(\la_{La}^{jk} \del I_{jk}^{La} +\hc\right) \n &
	+\frac{1}{m_T^2} \left( \la_T^{ij}\la_T^{ik\ast} (\del I^{Ta}_{ij})(\del I^{Ta\dag}_{ik}) +\frac{\la_T^{ii}\la_T^{jk\ast}}{2} (\del I^{Ta}_{ii})(\del I^{Ta\dag}_{jk}) +\hc \right) \\
\Rightarrow &
  \left(
    \frac{\la_0^{ii} \la_0^{jk}}{m_0^2}
   -\frac{\la_S^{ik} \la_S^{ij\ast}}{2m_S^2}
   -\frac{\la_L^{ii} \la_L^{jk} +\la_L^{ik} \la_L^{ij\ast}}{m_L^2}
	 -\frac{\la_T^{ik} \la_T^{ij\ast}}{2m_T^2}
  \right) O^H_{iijk} \n &
 +\left(
    \frac{\la_0^{ij} \la_0^{ik}}{m_0^2}
   -\frac{2\la_L^{ij} \la_L^{ik}}{m_L^2}
	 -\frac{\la_T^{jk} \la_T^{ii\ast}}{2m_T^2}
  \right) O^H_{ijik} \n &
 +\left(
    \frac{\la_0^{ij} \la_0^{ik\ast}}{m_0^2}
	 +\frac{\la_S^{ij} \la_S^{ik\ast}}{2m_S^2}
   -\frac{\la_L^{ij} \la_L^{ik\ast} +\la_L^{ii} \la_L^{jk\ast}}{m_L^2}
   -\frac{\la_T^{ij} \la_T^{ik\ast}}{2m_T^2}
  \right) O^H_{ijki}
\n &
 +\left(
   -\frac{\la_S^{ik} \la_S^{ij\ast}}{2m_S^2}
   +\frac{\la_L^{ik} \la_L^{ij\ast}}{m_L^2}
   -\frac{\la_T^{ik} \la_T^{ij\ast}}{2m_T^2}
  \right) O^T_{iijk}
 +\left(
    \frac{\la_L^{ij} \la_L^{ik}}{m_L^2}
	 -\frac{\la_T^{jk} \la_T^{ii\ast}}{2m_T^2}
  \right) O^T_{ijik} \n &
 +\left(
	  \frac{\la_S^{ij} \la_S^{ik\ast}}{2m_S^2}
   +\frac{\la_L^{ii} \la_L^{jk\ast}}{m_L^2}
   -\frac{\la_T^{ij} \la_T^{ik\ast}}{2m_T^2}
  \right) O^T_{ijki}
 +\hc.
\end{align}
These coefficients must be real to impose the custodial symmetry.
Additionally neglecting the imaginary part of the each three point scalar coupling, the requirement of the following relations guarantees the symmetry:
\begin{align}
 \frac{\la_T^{ii} \la_T^{jk}}{2m_T^2} =&
   \frac{\la_L^{ii} \la_L^{jk}}{m_L^2}
  +\frac{\la_S^{ij} \la_S^{ik}}{3m_S^2}, \\
 \frac{\la_T^{ij} \la_T^{ik}}{2m_T^2} =&
   \frac{\la_L^{ij} \la_L^{ik}}{m_L^2}
  -\frac{\la_S^{ij} \la_S^{ik}}{6m_S^2}.
\end{align}
They are trivially satisfied by the conditions discussed in the previous section.
Using the second condition, the relative signs of the flavor off diagonal couplings are fixed even if the flavor diagonal couplings do not exist.
\subsection{Type V}
The following effective Lagrangian is the type V derivative interactions given by the integration of the heavy scalars:
\begin{align}
\mcl{L}_\text{V}^S =&
  \frac{1}{m_0^2} \Bigl( 
	  \left( \la_0^{ij}(\del I^0_{ij}) +\hc \right)\left( \la_0^{kl} (\del I^0_{kl}) +\hc \right)
	 +\left( \la_0^{ik}(\del I^0_{ik}) +\hc \right)\left( \la_0^{jl} (\del I^0_{jl}) +\hc \right)
	 \n & \qquad
	 +\left( \la_0^{il}(\del I^0_{il}) +\hc \right)\left( \la_0^{jk} (\del I^0_{jk}) +\hc \right)
	\Bigr) \n &
 +\frac{1}{m_S^2} \Bigl( 
    \left(\la_S^{ij} \del I^S_{ij} \right) \left(\la_S^{kl\ast} \del I^{S\dag}_{kl} \right)
   +\left(\la_S^{ik} \del I^S_{ik} \right) \left(\la_S^{jl\ast} \del I^{S\dag}_{jl} \right)
   +\left(\la_S^{il} \del I^S_{il} \right) \left(\la_S^{jk\ast} \del I^{S\dag}_{jk} \right) +\hc
  \Bigr) \n &
 +\frac{1}{m_L^2} \Bigl( 
	  \left( \la_L^{ij}(\del I^{La}_{ij}) +\hc \right)\left( \la_L^{kl} (\del I^{La}_{kl}) +\hc \right)
	 +\left( \la_L^{ik}(\del I^{La}_{ik}) +\hc \right)\left( \la_L^{jl} (\del I^{La}_{jl}) +\hc \right)
	 \n & \qquad
	 +\left( \la_L^{il}(\del I^{La}_{il}) +\hc \right)\left( \la_L^{jk} (\del I^{La}_{jk}) +\hc \right)
  \Bigr) \n &
 +\frac{1}{m_T^2} \Bigl( 
    \left(\la_T^{ij} \del I^{Ta}_{ij} \right) \left(\la_T^{kl\ast} \del I^{Ta\dag}_{kl} \right)
   +\left(\la_T^{ik} \del I^{Ta}_{ik} \right) \left(\la_T^{jl\ast} \del I^{Ta\dag}_{jl} \right)
   +\left(\la_T^{il} \del I^{Ta}_{il} \right) \left(\la_T^{jk\ast} \del I^{Ta\dag}_{jk} \right) +\hc
  \Bigr) \\
\Rightarrow &
  \left(
    \frac{\la_0^{ij} \la_0^{kl}}{m_0^2}
   -\frac{\la_S^{jl} \la_S^{ik\ast}}{2m_S^2}
	-\frac{\la_L^{ij} \la_L^{kl} +\la_L^{il} \la_L^{jk\ast}}{m_L^2}
	-\frac{\la_T^{jl} \la_T^{ik\ast}}{2m_T^2}
  \right) O^H_{ijkl} \n &
 +\left(
    \frac{\la_0^{ij} \la_0^{kl\ast}}{m_0^2}
   -\frac{\la_S^{jk} \la_S^{il\ast}}{2m_S^2}
	-\frac{\la_L^{ij} \la_L^{kl\ast} +\la_L^{ik} \la_L^{jl\ast}}{m_L^2}
	-\frac{\la_T^{jk} \la_T^{il\ast}}{2m_T^2}
  \right) O^H_{ijlk} \n &
 +\left(
    \frac{\la_0^{ik} \la_0^{jl}}{m_0^2}
   -\frac{\la_S^{kl} \la_S^{ij\ast}}{2m_S^2}
	-\frac{\la_L^{ik} \la_L^{jl} +\la_L^{il} \la_L^{jk}}{m_L^2}
	-\frac{\la_T^{kl} \la_T^{ij\ast}}{2m_T^2}
  \right) O^H_{ikjl} \n &
 +\left(
    \frac{\la_0^{ik} \la_0^{jl\ast}}{m_0^2}
   +\frac{\la_S^{jk} \la_S^{il\ast}}{2m_S^2}
	-\frac{\la_L^{ij} \la_L^{kl\ast} +\la_L^{ik} \la_L^{jl\ast}}{m_L^2}
   -\frac{\la_T^{jk} \la_T^{il\ast}}{2m_T^2}
  \right)O^H_{iklj} \n &
 +\left(
    \frac{\la_0^{il} \la_0^{jk}}{m_0^2}
   +\frac{\la_S^{kl} \la_S^{ij\ast}}{2m_S^2}
	-\frac{\la_L^{ik} \la_L^{jl} +\la_L^{il} \la_L^{jk}}{m_L^2}
	-\frac{\la_T^{kl} \la_T^{ij\ast}}{2m_T^2}
  \right) O^H_{iljk} \n &
 +\left(
    \frac{\la_0^{il} \la_0^{jk\ast}}{m_0^2}
   +\frac{\la_S^{jl} \la_S^{ik\ast}}{2m_S^2}
	-\frac{\la_L^{ij} \la_L^{kl} +\la_L^{il} \la_L^{jk\ast}}{m_L^2}
	-\frac{\la_T^{jl} \la_T^{ik\ast}}{2m_T^2}
  \right) O^H_{ilkj}
\n &
 +\left(
   -\frac{\la_S^{jl} \la_S^{ik\ast}}{2m_S^2}
   +\frac{\la_L^{il} \la_L^{jk\ast}}{m_L^2}
	-\frac{\la_T^{jl} \la_T^{ik\ast}}{2m_T^2}
  \right) O^T_{ijkl}
 +\left(
   -\frac{\la_S^{jk} \la_S^{il\ast}}{2m_S^2}
   +\frac{\la_L^{ik} \la_L^{jl\ast}}{m_L^2}
	-\frac{\la_T^{jk} \la_T^{il\ast}}{2m_T^2}
  \right) O^T_{ijlk} \n &
 +\left(
   -\frac{\la_S^{kl} \la_S^{ij\ast}}{2m_S^2}
   +\frac{\la_L^{il} \la_L^{jk}}{m_L^2}
	-\frac{\la_T^{kl} \la_T^{ij\ast}}{2m_T^2}
  \right) O^T_{ikjl}
 +\left(
	 \frac{\la_S^{jk} \la_S^{il\ast}}{2m_S^2}
   +\frac{\la_L^{ij} \la_L^{kl\ast}}{m_L^2}
   -\frac{\la_T^{jk} \la_T^{il\ast}}{2m_T^2}
  \right) O^T_{iklj} \n &
 +\left(
	 \frac{\la_S^{kl} \la_S^{ij\ast}}{2m_S^2}
   +\frac{\la_L^{ik} \la_L^{jl}}{m_L^2}
   -\frac{\la_T^{kl} \la_T^{ij\ast}}{2m_T^2}
  \right) O^T_{iljk}
 +\left(
	 \frac{\la_S^{jl} \la_S^{ik\ast}}{2m_S^2}
   +\frac{\la_L^{ij} \la_L^{kl}}{m_L^2}
   -\frac{\la_T^{jl} \la_T^{ik\ast}}{2m_T^2}
  \right) O^T_{ilkj}
 +\hc .
\end{align}
Assuming that all of couplings are real, the following conditions impose the custodial symmetry on the above effective Lagrangian:
\begin{align}
 0 =&
   -\la_S^{ik}\la_S^{jl} +\la_S^{il}\la_S^{jk} +\la_S^{ij}\la_S^{kl}, \\
 \frac{\la_T^{ij} \la_T^{kl}}{2m_T^2} =&
   \frac{\la_L^{ij} \la_L^{kl}}{m_L^2}
  +\frac{\la_S^{ik} \la_S^{jl} +\la_S^{il} \la_S^{jk}}{6m_S^2}, \\
 \frac{\la_T^{ik} \la_T^{jl}}{m_T^2} =&
   \frac{\la_L^{ik} \la_L^{jl}}{m_L^2}
  +\frac{\la_S^{ij} \la_S^{kl} -\la_S^{il} \la_S^{jk}}{6m_S^2}, \\
 \frac{\la_T^{il} \la_T^{jk}}{m_T^2} =&
   \frac{\la_L^{il} \la_L^{jk}}{m_L^2}
  -\frac{\la_S^{ij} \la_S^{kl} +\la_S^{ik} \la_S^{jl}}{6m_S^2} .
\end{align}
In addition to the three conditions like those appearing in the type III and IV cases, a relation among the $\textbf{1}_1$ couplings is required.
However, these conditions are weaker than the conditions which have been discussed in the type III.
\subsection{Gauge currents}
The heavy vector fields of the $\textbf{3}_0$ and the $\textbf{1}_0$ representations can couple with the SM gauge currents.
After integrating out, their couplings appear in contributions to the oblique parameters.
Expanding the related part of Eq.~\eqref{EqVec}, we obtain
\begin{align}
 \mcl{L}_\text{eff}^V \supset &
  \sum_i \rh_0 \frac{g' g_0^{ii}}{m_0^2} B^{\mu \nu} \left(
	  i(D_\mu H)^\dag (D_\nu H) -i(D_\nu H)^\dag (D_\mu H)
	 -g W^a_{\mu\nu} H^\dag \si^a H -g' B_{\mu \nu} H^\dag H
  \right) \n &
 +\sum_i \rh_L \frac{g g_L^{ii}}{m_L^2} W^{a\mu \nu} \left(
    i(D_\mu H)^\dag \si^a (D_\nu H) -i(D_\nu H)^\dag \si^a (D_\mu H)
	 -g W^a_{\mu\nu} H^\dag H -g' B_{\mu \nu} H^\dag \si^a H 
  \right) \n &
  -\frac{1}{2} \left( \frac{g' \rh_0}{m_0} \right)^2 
   (\del \cdot B)\cdot (\del \cdot B)
  -\frac{1}{2} \left( \frac{g \rh_L}{m_L} \right)^2 
   (D\cdot W)^a \cdot (D\cdot W)^a .
\end{align}
Following the definition of Ref.~\cite{RefEWPM}, the parameters are
\begin{align}
 \hat{S} =&
   -g^2 \biggl(
	   \rh_0 \left(
		  \sum_i g_0^{ii} \frac{v_i^2}{m_0^2}
     \right)
	  +\rh_L \left(
		  \sum_i g_L^{ii} \frac{v_i^2}{m_L^2}
     \right)
	 \biggr), \\
 W =& \frac{1}{2} \left( \frac{g  \rh_L m_W}{m_L} \right)^2, \\
 Y =& \frac{1}{2} \left( \frac{g' \rh_0 m_W}{m_0} \right)^2.
\end{align}
The parameter $\hat{S}$ is modified by the effect of the additional doublets.
The others are the same as the models including only the SM Higgs doublet.

The strongest constraint comes from $\hat{S}$, and its contribution can be written
\begin{align}
 \hat{S}
  =&
	 -4m_W^2 \left(
	   \frac{\rh_0}{m_0^2} \sqrt{\sum_i g_0^{ii\, 2}}\, \hat{\xi}_0
		+\frac{\rh_L}{m_L^2} \sqrt{\sum_i g_L^{ii\, 2}}\, \hat{\xi}_L
	 \right) \cdot \hat{\mu}, \\
 \in & 
   4m_W^2 \sqrt{\sum_i \left( \left( \frac{\rh_0 g_0^{ii}}{m_0^2} \right)^2 +\left( \frac{\rh_L g_L^{ii}}{m_L^2} \right)^2 \right)}\, [-1,1] ,
\end{align}
where $\hat{\xi}_{0,L}$ and $\hat{\mu}$ are $N$ dimensional unit vectors defined by
\begin{align}
 (\hat{\xi}_{0,L})_i =& \frac{g_{0,L}^{ii}}{\sqrt{ \sum g_{0,L}^{ii\,2} }}, \\
 (\hat{\mu})_i =& \frac{v_i^2}{v^2}.
\end{align}

As is well known, the absolute value of $\hat{S}$ should be smaller than about 0.001.
For example, in the one Higgs doublet model, assuming that $m_0 = m_L = M$, $g_X = 1$, and $\rh_X \sim m_W/m_X$, the constraint implies $M \gtrsim 1.6\TeV$.
\section{Properties of the given derivative interactions}
\label{SecProp}
In this section, we show several differences between the effective Lagrangian of the general Higgs derivative interactions and that of the case generated by integrating out which we have calculated in the previous section.

First, we discuss the DOFs of the Wilson coefficients.
If the DOFs of the general one are larger than of the other one, we find that UV models are described by a composite Higgs model because the DOFs of composite Higgs models can be the same as the general one; see Ref.~\cite{RefKy}.

Secondly, we explicitly show the derivative interactions in the usual two Higgs doublet model and the inert doublet model.
Several properties appearing in the vector boson scatterings\footnotemark are discussed in each model.
\footnotetext{
 Strictly speaking, they are not the vector boson scatterings but the vector and scalar boson scatterings because additional doublets give us additional scalar particles not eaten by vector bosons.
 However, for simplicity, we call them vector boson scatterings.
}
\subsection{Degrees of freedom}
We compare the DOFs of the Wilson coefficients between the effective Lagrangian given by integrating out and the general one.
Naively, the DOFs of the general effective Lagrangian for the derivative interactions are proportional to $N^4$, while those given by integrating out are proportional to $N^2$.
Hence, the DOFs of the former one can be much larger than of the latter one for the large $N$ case.

Following Ref.~\cite{RefKoy}, the DOFs of the general effective theory with $n$ $Z_2$-odd doublets\footnotemark are calculated as Table~\ref{TabGendof}.
\footnotetext{
 This discrete symmetry is introduced to sequester a part of doublets from the Standard Model Higgs.
 The minimal example is the inert doublet model briefly studied later.
}
\begin{table}[t]
\centering
\begin{tabular}{c|l}
 General & DOFs \\ \hline
 Re  & $(N^2 (N^2 +3) -4n(N-n)(N^2 -2nN  +2n^2))/2$ \\
 Im  & $(N^2 (N^2 -1) -4n(N-n)(N^2 -2nN  +2n^2))/2$ \\
 Cust.  & $(N^2 (N^2 +2) -2n(N-n)(2N^2-1 -4n(N-n)))/3$
\end{tabular}
\caption{
The DOFs of the general effective Lagrangian.
The first and second ones are respectively the real and the imaginary DOFs in the case including $n$ $Z_2$-odd doublets.
Imposing the custodial symmetry, the real one becomes the third one, and the imaginary one vanishes.
}
\label{TabGendof}
\end{table}
The DOFs given by integrating out the $Z_2$-even scalars, the $Z_2$-odd scalars, and the vectors are shown in Tables~\ref{TabEvens}, \ref{TabOdds}, and \ref{TabVec}, respectively.
\begin{table}[t]
\centering
\begin{tabular}{c|lll}
 $Z_2$-even & Re & Im ($N\geq 2$) & Sum \\ \hline
 $\textbf{1}_0$ & $(N(N+1)-2n(N-n))/2$ & $(N(N-1)-2n(N-n))/2$     & $N^2-2n(N-n)$      \\
 $\textbf{1}_1$ & $(N(N-1)-2n(N-n))/2$ & $((N+1)(N-2)-2n(N-n))/2$ & $N(N-1)-2n(N-n)-1$ \\
 $\textbf{3}_0$ & $(N(N+1)-2n(N-n))/2$ & $(N(N-1)-2n(N-n))/2$     & $N^2-2n(N-n)$      \\
 $\textbf{3}_1$ & $(N(N+1)-2n(N-n))/2$ & $((N+2)(N-1)-2n(N-n))/2$ & $N(N+1)-2n(N-n)-1$ \\
 Sum            & $N(2N+1)-4n(N-n)$    & $N(2N-1)-4n(N-n)-2$      & $2(2N^2-1-4n(N-n))$
\end{tabular}
\caption{
 The DOFs generated by integrating out the heavy $Z_2$-even scalars with $n$ $Z_2$-odd doublets.
}
\label{TabEvens}
\end{table}
\begin{table}[t]
\centering
\begin{tabular}{c|lll}
 $Z_2$-odd & Re & Im ($N\geq 2$) & Sum \\ \hline
 $\textbf{1}_0$ & $(N-n)n$ & $(N-n)n$   & $2(N-n)n$   \\
 $\textbf{1}_1$ & $(N-n)n$ & $(N-n)n-1$ & $2(N-n)n-1$ \\
 $\textbf{3}_0$ & $(N-n)n$ & $(N-n)n$   & $2(N-n)n$   \\
 $\textbf{3}_1$ & $(N-n)n$ & $(N-n)n-1$ & $2(N-n)n-1$ \\
 Sum            & $4(N-n)n$& $4(N-n)n-2$& $8(N-n)n-2$
\end{tabular}
\caption{
 The DOFs generated by integrating out the heavy $Z_2$-odd scalar fields with $n$ $Z_2$-odd doublets.
}
\label{TabOdds}
\end{table}
\begin{table}[t]
\centering
\begin{tabular}{c|lll}
 Vector & Re & Im ($\geq 2$) & Sum \\ \hline
 $\textbf{1}_0$ & $N$  & $0$   & $N$    \\
 $\textbf{1}_1$ & $N$  & $N-1$ & $2N-1$ \\
 $\textbf{3}_0$ & $N$  & $0$   & $N$    \\
 Sum            & $3N$ & $0$   & $4N-1$
\end{tabular}
\caption{
 The DOFs generated by integrating out the heavy vector fields.
}
\label{TabVec}
\end{table}
The total DOFs of the heavy scalars are independent of the number of the $Z_2$-odd doublets.

For simplicity, we assume all couplings are real.
Then, as discussed in the previous section, the DOFs of the derivative interactions are the same as with the sum of the real couplings in the $\textbf{1}_0$ and the $\textbf{3}_0$ representations.
The $\textbf{3}_1$ scalar couplings and the $\textbf{1}_1$ vector couplings are fixed to preserve the custodial symmetry, and the contributions of the $\textbf{1}_1$ scalar disappear.

The DOFs of the general effective Lagrangian with the custodial symmetry are minimized if half of the doublets are $Z_2$-odd, and the number of DOFs is
\begin{align}
  \frac{1}{6} N^2 (N^2 +5).
\end{align}
Even if $N$ is an odd number, this minimum number of DOFs is obtained for $n=(N\pm 1)/2$.
The total DOFs generated by integrating out is
\begin{align}
 N(N+3).
\end{align}
Hence, the difference of these DOFs between the general one and the case of integrating out is
\begin{align}
 \frac{1}{6} N(N^3 -N -18).
\end{align}

If $N \geq 3$, the maximal number of DOFs of the case of integrating out is smaller than that of the other one.
Their difference is three for $N=3$.
Assuming the vector couplings are unique, like the gauge theory, for each representation, the difference becomes larger.
Without using the assumption to simplify, the difference of the DOFs is occasionally the same as these results.
The difference between these DOFs becomes larger and larger if $N$ increases.
Therefore, the Higgs derivative interactions generated by integrating out is a proper subset of the general one.

If we do not assume the alignment limit of the Georgi--Machacek type structure, the triplet scalar couplings are inefficient.
The DOFs of this case is
\begin{align}
 \frac{N(N+5)}{2}.
\end{align}
$N\geq 3$ is required when the maximum number of effective DOFs given by the case of integrating out is smaller than that given by the general one, namely the maximum DOFs of the composite one.
\subsection{Two Higgs doublet models}
The explicit forms of the derivative interactions and the contributions to some vector boson scattering processes are shown below on a couple of two Higgs doublet models.
One of them includes an additional doublet which has the vacuum expectation value, and the other includes an inert doublet. 
We discuss only the processes of the initial states which consist of the longitudinal $W$ boson pairs, since their cross sections are typically larger than the others.

In the following study, we neglect the imaginary parts of the couplings and separately apply the custodial symmetry on the heavy scalars and the heavy vectors.
If both of them simultaneously  exist in the higher scale, the extra contribution violating the symmetry can be parametrically canceled as a fine tuning.

The current constraints of these models without the higher dimensional operators have been studied by Ref.~\cite{Ref2HDM,RefIdm1,RefIdm2}.

\subsubsection{With an additional Higgs doublet}
Including the second Higgs doublet without the $Z_2$ symmetry, the $Z_2$-even scalars and the vectors can be introduced as the sources of the derivative interactions.

Integrating out the heavy scalars, the following derivative interactions are obtained:
\begin{align}
 \mcl{L}_\text{Scalar} =&
  \left(
    \frac{ \la_0^{11\,2}}{m_0^2}
   -\frac{3\la_L^{11\,2}}{m_L^2}
  \right) O^H_{1111}
 +\left(
    \frac{\la_0^{11} \la_0^{12}}{m_0^2} -\frac{3\la_L^{11} \la_L^{12}}{m_L^2}
  \right) ( O^H_{1112} +\hc ) \n &
  +\left(
    \frac{ \la_0^{11} \la_0^{22}}{m_0^2}
   -\frac{2\la_L^{12\,2} +\la_L^{11} \la_L^{22}}{m_L^2}
  \right) O^H_{1122}
 +\left(
    \frac{ \la_0^{12\,2}}{m_0^2}
   -\frac{2\la_L^{12\,2} +\la_L^{11} \la_L^{22}}{m_L^2}
  \right) O^H_{1221} \n &
 +\left(
    \frac{ \la_0^{12\,2}}{m_0^2}
   -\frac{2\la_L^{12\,2} +\la_L^{11} \la_L^{22}}{m_L^2}
  \right) ( O^H_{1212} +\hc )
 +\left(
   \frac{\la_0^{12} \la_0^{22}}{m_0^2} -\frac{3\la_L^{12} \la_L^{22}}{m_L^2}
  \right) ( O^H_{2221} +\hc ) \n &
 +\left(
   \frac{ \la_0^{22\,2}}{ m_0^2}
  -\frac{3\la_L^{22\,2}}{ m_L^2}
 \right) O^H_{2222} \n &
 -\left(
    \frac{\la_L^{12\,2} -\la_L^{11} \la_L^{22}}{m_L^2}
  \right) O^T_{1221}
 +\left(
    \frac{\la_L^{12\,2} -\la_L^{11} \la_L^{22}}{m_L^2}
  \right) ( O^T_{1212} +\hc ),
\end{align}
where seven couplings are included in the Wilson coefficients.

Integrating out heavy vectors, the following effective Lagrangian is obtained with four couplings:
\begin{align}
 \mcl{L}_\text{Vector} =&
  3\left(
    \frac{g_0^{11\,2}}{m_0^2}
   +\frac{g_L^{11\,2}}{m_L^2}
  \right) O^H_{1111}
 +3\left(
    \frac{g_L^{11} g_L^{22}}{m_L^2}
  \right) O^H_{1221} \n &
 +3\left(
    \frac{g_0^{11} g_0^{22}}{m_0^2}
  \right) ( O^H_{1212} +\hc )
 +3\left(
   \frac{g_0^{22\,2}}{m_0^2}
  +\frac{g_L^{22\,2}}{m_L^2}
 \right) O^H_{2222} \n &
 +\left(
    \frac{g_0^{11} g_0^{22}}{m_0^2}
   -\frac{g_L^{11} g_L^{22}}{m_L^2}
  \right) O^T_{1122}
 +\left(
   \frac{g_L^{11} g_L^{22}}{m_L^2}
  \right) O^T_{1221}
 -\left(
   \frac{g_0^{11} g_0^{22}}{m_0^2}
  \right) ( O^T_{1212} +\hc ).
\label{EqVecCoeff}
\end{align}
If the above interactions dominate the vector boson scatterings, the collision energies of the vector bosons are much larger than their masses.
Hence we neglect the masses and consider the $W$ boson pairs as the initial states of the scatterings in the following part.
The cross sections of the other initial states are much smaller than this state.
All of the cross sections have been given by Ref.~\cite{RefKoy} in terms of the Wilson coefficients.
We follow its notation.
For the leading order, the mixing matrices are defined as
\begin{align}
 \begin{pmatrix} h \\ H \end{pmatrix} =&
 \begin{pmatrix} c_\al & s_\al \\ -s_\al & c_\al \end{pmatrix}
 \begin{pmatrix} S_1 \\ S_2 \end{pmatrix}, \\
 \begin{pmatrix} W_L^\pm \\ H^\pm \end{pmatrix} =&
 \begin{pmatrix} c_\be & s_\be \\ -s_\be & c_\be \end{pmatrix}
 \begin{pmatrix} C_1^\pm \\ C_2^\pm \end{pmatrix},
\end{align}
where $h$, $H$, $W_L^\pm$, and $H^\pm$ are, respectively, the SM-like Higgs, the heavy Higgs, the Goldstone boson eaten by $W^\pm$ and the charged Higgs. 
The fields given as  $S_i$ and $C_i^\pm$ are the scalar and the charged scalar components of the doublets, $H_i$.

In the one Higgs doublet case, the derivative interactions are effectively written by a certain parameter.
Then the ratios of the cross sections are independent of the model parameters.
This feature is not preserved in the two Higgs doublet models because of two mixing angles, $\al$ and $\be$.
However, in the decoupling limit, which is favored by current experimental results, the feature is recovered.
The amplitudes of the vector boson scatterings are expressed by a certain coefficient like the one Higgs case for processes appearing in the case.
Therefore, these processes are insensitive to the two Higgs nature.

Considering the processes including an additional Higgs boson, the same sign $W$ boson scattering with a charged Higgs boson is a promising process.
The cross section of the sub process is 
\begin{align}
 \si (W_L^\pm W_L^\pm \to W_L^\pm H^\pm) =
 \frac{\hat{s}}{16\pi} \left( C_{S0}(\be) +C_{SL}(\be) +C_{V0}(\be) +C_{VL}(\be) \right)^2,
\end{align}
with
\begin{align}
 C_{S0}(\be) = \frac{1}{8 m_0^2} \Bigl( &
   -(2s_{2\be}+s_{4\be}) \la_0^{11\,2}
	 +4(c_{2\be}+c_{4\be}) \la_0^{11} \la_0^{12}
	 +2s_{4\be} ( \la_0^{11} \la_0^{22} +2\la_0^{12\,2} )  \n &
	 +4(c_{2\be}-c_{4\be}) \la_0^{12} \la_0^{22}
	 +(2s_{2\be}-s_{4\be}) \la_0^{22\,2}
	\Bigr),\\
 C_{SL}(\be) = \frac{3}{8 m_L^2} \Bigl( &
    (2s_{2\be}+s_{4\be}) \la_L^{11\,2}
	 -4(c_{2\be}+c_{4\be}) \la_L^{11} \la_L^{12}
	 -2s_{4\be} ( 2\la_L^{12\,2} +\la_L^{11} \la_L^{22} ) \n &
	 -4(c_{2\be}-c_{4\be}) \la_L^{12} \la_L^{22}
	 -(2s_{2\be}-s_{4\be}) \la_L^{22\,2}
	\Bigr),\\
 C_{V0}(\be) = \frac{3}{8 m_0^2} \Bigl( &
   -(2s_{2\be}+s_{4\be}) g_0^{11\,2}
	 +2s_{4\be} g_0^{11} g_0^{22}
	 +(2s_{2\be}-s_{4\be}) g_0^{22\,2}
	\Bigr),\\
 C_{VL}(\be) = \frac{3}{8 m_L^2} \Bigl( &
   -(2s_{2\be}+s_{4\be}) g_L^{11\,2}
	 +2s_{4\be} g_L^{11} g_L^{22}
	 +(2s_{2\be}-s_{4\be}) g_L^{22\,2}
	\Bigr),
\end{align}
where $\hat{s}$ is the squared collision energy of sub processes.

If the flavor off diagonal couplings, $\la_{\cdots}^{12}$, are negligibly small, the $\be$ dependences of the scalar contributions are the same as the vector ones.
Assuming all couplings are the same, the scalar contributions in the amplitude proportional to $2c_{2\be}+s_{4\be}$, which vanishes at $\be=\pi/4$.
The vector contributions disappear for any $\be$ in this case.
Different values of $\be$ pick up different couplings.
The contributions by the heavy vectors do not include the coefficients of $O^H_{1112}$ and $O^H_{2221}$.
Hence, the effects of integrating out the heavy vectors are suppressed when $\be \sim 0$, $\pi/2$ or $\pi$.

The cross section including two charged Higgs bosons in the final state is given as follows:
\begin{align}
 \si (W_L^\pm W_L^\pm \to H^\pm H^\pm) =
  \frac{\hat{s}}{32\pi} \left( C_{S0}(\be) +C_{SL}(\be) +C_{V0}(\be) +C_{VL}(\be) \right)^2,
\end{align}
where
\begin{align}
  C_{S0}(\be) = \frac{1}{8m_0^2} \Bigl( &
	 (1 -c_{4\be}) \left( \la_0^{11\,2} -4\la_0^{12\,2} +\la_0^{22\,2} \right)
	-4s_{4\be} (\la_0^{11} \la_0^{12} -\la_0^{12} \la_0^{22}) 
	+2(3 +c_{4\be}) \la_0^{11} \la_0^{22}
  \Bigr), \\
  C_{SL}(\be) = \frac{1}{8m_L^2} \Bigl( &
	 (1 -c_{4\be})
    \left( -3\la_L^{11\,2} -3\la_L^{22\,2} +8\la_L^{12\,2} +4\la_L^{11} \la_L^{22} \right)
	+12s_{4\be} \left( \la_L^{11} \la_L^{12} -\la_L^{12} \la_L^{22} \right) \n &
	-2(3 +c_{4\be}) \left( 2\la_L^{12\,2} +\la_L^{11} \la_L^{22} \right)
  \Bigr), \\
  C_{V0}(\be) = \frac{3}{8m_0^2} \Bigl( &
	 (1 -c_{4\be}) \left( g_0^{11} -g_0^{22} \right)^2
  \Bigr), \\
  C_{VL}(\be) = \frac{3}{8m_L^2} \Bigl( &
	 (1 -c_{4\be}) \left( g_L^{11} -g_L^{22} \right)^2
  \Bigr).
\end{align}
Assuming, again, that all couplings are the same, the scalar contributions disappear at $\be=\pi/4$ again.
The vector contributions also disappear for any $\be$ like the previous one.

The above two processes are affected by the couplings between the heavy bosons and the additional Higgs doublet.
However, in the unique coupling limit, they simultaneously vanish at $\be =\pi /4$.
To see the $\be$ dependent behavior in this limit, we show the cross section of the same sign $W_L$ scattering,
\begin{align}
 \si (W_L^\pm W_L^\pm \to W_L^\pm W_L^\pm) =
  \frac{\hat{s}}{32\pi} \left( C_{S0}(\be) +C_{SL}(\be) +C_{V0}(\be) +C_{VL}(\be) \right)^2,
	\label{EqSsww}
\end{align}
where
\begin{align}
  C_{S0}(\be) = \frac{1}{8m_0^2} \Bigl( &
	 (3 +4c_{2\be} +c_{4\be}) \la_0^{11\,2}
	+4(2s_{2\be} +s_{4\be}) \la_0^{11} \la_0^{12}
	+2(1 -c_{4\be}) \left(\la_0^{11} \la_0^{22} +2\la_0^{12\,2} \right) \n &
	+4(2s_{2\be} -s_{4\be}) \la_0^{12} \la_0^{22}
	+(3 -4c_{2\be} +c_{4\be}) \la_0^{22\,2}
  \Bigr), \\ 
  C_{SL}(\be) = \frac{3}{8m_L^2} \Bigl( &
	-(3 +4c_{2\be} +c_{4\be}) \la_L^{11\,2}
	-4(2s_{2\be} +s_{4\be}) \la_L^{11} \la_L^{12}
	-2(1 -c_{4\be}) \left( 2\la_L^{12\,2} +\la_L^{11} \la_L^{22} \right) \n &
	-4(2s_{2\be} -s_{4\be}) \la_L^{12} \la_L^{22}
	-(3 -4c_{2\be} +c_{4\be}) \la_L^{22\,2}
  \Bigr), \\
  C_{V0}(\be) = \frac{3}{8m_0^2} \Bigl( &
	 (3 +4c_{2\be} +c_{4\be}) g_0^{11\,2}
	+2(1 -c_{4\be}) g_0^{11} g_0^{22}
	+(3 -4c_{2\be} +c_{4\be}) g_0^{22\,2}
  \Bigr), \\
  C_{VL}(\be) = \frac{3}{8m_L^2} \Bigl( &
	 (3 +4c_{2\be} +c_{4\be}) g_L^{11\,2}
	+2(1 -c_{4\be}) g_L^{11} g_L^{22}
	+(3 -4c_{2\be} +c_{4\be})g_L^{22\,2}
  \Bigr).
\end{align}
This mode is a promising mode to see the Higgs derivative interactions in the one Higgs doublet models.
As we have mentioned, it does not reflect the two Higgs nature in the decoupling limit.
However, in the unique coupling limit, its mixing dependence is different from the previous ones.
For the scalar contributions, the cross section of the sub process is maximized at $\be=\pi/4$ where the cross sections of the previous two processes disappear.
The vector contributions are independent of $\be$ and also do not vanish.

The vector boson scatterings with the SM particles are not sensitive to the two Higgs nature in the decoupling limit.
However, comparing them with other scatterings including additional Higgs bosons help us to see the couplings between the doublets and the other heavy bosons behind them.

According to Ref.~\cite{RefIlc}, the Wilson coefficients, $C$, which stands for the functions in the parentheses of Eq.~\eqref{EqSsww}, are constrained thus:
\begin{align}
 \frac{1}{\sqrt{C}} \gtrsim  2.5\TeV,
\end{align}
after the running of ILC 500\TeV  with 500~fb$^{-1}$.
The $\textbf{3}_0$ scalar and the vector contributions are numerically enhanced comparing with the $\textbf{1}_0$ scalar contribution.
They might be sensitive to the physics of even a bit higher scale.
\subsubsection{With inert doublet}
Another possibility of the two doublet scenario is the inert doublet model, where the additional doublet is a $Z_2$-odd scalar and does not possess the vacuum expectation value.
In this model, physical states also respect the discrete symmetry and no mixing occurs between the doublets.

We can introduce the $Z_2$-even scalars, the $Z_2$-odd scalars, and the vectors.
The following effective Lagrangians are obtained by integrating out:
\begin{align}
 \mcl{L}_\text{Seven} =&
  \left(
    \frac{ \la_0^{11\,2}}{m_0^2}
   -\frac{3\la_L^{11\,2}}{m_L^2}
  \right) O^H_{1111}
  +\left(
    \frac{\la_0^{11} \la_0^{22}}{m_0^2}
   -\frac{\la_L^{11} \la_L^{22}}{m_L^2}
  \right) O^H_{1122}
 -\left(
   \frac{\la_L^{11} \la_L^{22}}{m_L^2}
  \right) O^H_{1221} \n &
 -\left(
    \frac{ \la_L^{11} \la_L^{22}}{ m_L^2}
  \right) ( O^H_{1212} +\hc )
 +\left(
   \frac{ \la_0^{22\,2}}{ m_0^2}
  -\frac{3\la_L^{22\,2}}{ m_L^2}
 \right) O^H_{2222} \n &
 +\left(
   \frac{ \la_L^{11} \la_L^{22}}{m_L^2}
  \right) O^T_{1221}
 -\left(
	  \frac{\la_L^{11} \la_L^{22}}{m_L^2}
  \right) ( O^T_{1212} +\hc ),
\end{align}
and
\begin{align}
 \mcl{L}_\text{Sodd} =&
  -\left(
    \frac{2\la_L^{12\,2}}{m_L^2}
  \right) O^H_{1122}
 +\left(
    \frac{ \la_0^{12\,2}}{m_0^2}
   -\frac{2\la_L^{12\,2}}{m_L^2}
  \right) O^H_{1221}
 +\left(
    \frac{ \la_0^{12\,2}}{m_0^2}
   -\frac{2\la_L^{12\,2}}{m_L^2}
  \right) ( O^H_{1212} +\hc ) \n &
 -\left(
    \frac{ \la_L^{12\,2}}{m_L^2}
  \right) O^T_{1221}
 +\left(
    \frac{\la_L^{12\,2}}{m_L^2}
  \right) ( O^T_{1212} +\hc ).
\end{align}
The flavor diagonal and the off diagonal couplings respectively originate from the $Z_2$-even and the $Z_2$-odd scalars.
The heavy vector contributions are the same as Eq.~\eqref{EqVecCoeff} because their couplings are diagonal for the Higgs indices.

Since doublets do not mix with each other, the processes consisting of the SM particles are the same as the one Higgs doublet case studied in Ref.~\cite{RefLrv}.
At least two new scalars are required in the effective four point vertices to see some new effects of the above effective operators.

A fascinating scenario of the inert doublet model is that the additional CP-even neutral component becomes the dark matter.
According to Ref.~\cite{RefIdm1}, the current experimental constraints divide the parameter region into two parts.
One is the light dark matter scenario, where the dark matter is about 90\GeV, and the other is the heavy dark matter scenario, where the dark matter is heavier than 500\GeV.
If the light dark matter exists, the dark matter pair production cross section by the vector boson scattering can be large at the high energy region.
For example, the cross section of this sub process is
\begin{align}
  \si(W_L^+ W_L^- \to H H)
 =\frac{\hat{s}}{32\pi} \left(
   \frac{\la_0^{11} \la_0^{22}}{m_0^2}
	-\frac{\la_L^{11} \la_L^{22} +2\la_L^{12\,2}}{m_L^2}
 \right)^2.
\end{align}
This contribution can enhance the process of two forward jets plus large missing energy at high energy.
The heavy vector contributions do not appear in this process.
The cross section of $W_L^\pm W_L^\pm \to H^\pm H^\pm$ is proportional to the above process, so that this mode is also insensitive to the effects of the heavy vector bosons.
Then the pair production of the charged Higgs bosons help us to see the vector couplings,
\begin{align}
  \si(W_L^+ W_L^- \to H^+ H^-)
 =\frac{\hat{s}}{48\pi} \biggl( &
   \left(
	   \frac{3g_L^{11}g_L^{22}}{m_L^2}
		+\frac{-3\la_0^{11} \la_0^{22} +2\la_0^{12\,2}}{2m_0^2}
		+\frac{\la_L^{11}\la_L^{22} +2\la_L^{12\,2}}{2m_L^2}
	 \right)^2 \n &
  +\frac{3}{4}\left(
     \frac{\la_0^{11}\la_0^{22}}{m_0^2}
	  -\frac{\la_L^{11}\la_L^{22} +2\la_L^{12\,2}}{m_L^2}
  \right)^2
 \biggr).
\end{align}
Even using this mode, the coupling with the $\textbf{1}_0$ vector does not appear.
Its contribution appear in $W_L^+ W_L^- \to H A$.
However, the cross section of this mode is about one order of magnitude smaller than the above processes.
\section{Conclusion}
\label{SecCon}
The Higgs derivative interactions are important to study the UV structure of the Higgs sector and the origin of the EWSB.
They are generated by strongly interacting models and heavy particle integrations.

We have studied these derivative interactions generated by integrating out in the models including any number of the Higgs doublets.
The three point couplings between the heavy bosons and the Higgs bilinears give us the tree level contribution to the derivative interactions.
These effects are expected to appear in the vector boson scattering processes.

The DOFs of the Wilson coefficients in the general effective Lagrangian of these interactions increase as $N^4$, while those given by integrating out increase as $N^2$.
Hence, a part of the parameter region in the general effective Lagrangian cannot be described with the effective Lagrangian given by the heavy particle integrations.
If the integration cannot generate the given DOFs, the UV completion has to be a composite Higgs model.
Imposing the custodial symmetry, we found that, if $N\geq 3$, the DOFs of the general one can be larger than those of the other one. 

We have also investigated the effects of the derivative interactions to the vector boson scatterings in the usual two Higgs doublet model and the inert doublet model with the assumptions explained there.

For the two Higgs doublet model, we have assumed the decoupling limit.
In this limit, at least, one new Higgs boson is required to see the couplings between the heavy bosons and the additional Higgs doublet.
The cross sections of $W_L^\pm W_L^\pm \to W_L^\pm H^\pm$ and $W_L^\pm W_L^\pm \to H^\pm H^\pm$ are shown there.
If we assume all couplings are the same, like the gauge theory, they can simultaneously disappear at a certain $\be$.
However, the cross section of $W_L^\pm W_L^\pm \to W_L^\pm W_L^\pm$ does not disappear even in these cases.
Therefore, the usual vector boson scatterings are also important to investigate the derivative interactions in the two Higgs doublet models.

For the case of the inert doublet, two new Higgs bosons are required to observe new effects.
If the additional CP-even Higgs boson is the dark matter, the two jet and large missing transverse momentum event is affected by the derivative interactions through $W_L^+ W_L^- \to H H$.
This process is generated by only the heavy scalar contributions.
Another promising process, $W_L^\pm W_L^\pm \to H^\pm H^\pm$, is also independent of the heavy vector contributions.
The $\textbf{3}_0$ vector contribution appears in $W_L^+ W_L^- \to H^+ H^-$.
To see the contribution of the $\textbf{1}_0$ vector, we need to observe $W_L^+ W_L^- \to H A$, whose cross section is much smaller than the others.
Therefore, in the inert doublet scenario, we have found that it is difficult to find the effects of the heavy vector contributions through the Higgs derivative interactions.
\section*{Acknowledgement}
The authors thank M. Tanabashi for useful comments.
The work of Y.K. was supported in part by Grant-in-Aid for JSPS Fellows, Japan Society for the Promotion of Science (JSPS), No.25.2309.
The work of Y.Y. has been supported in part by the Ministry of Economy and Competitiveness (MINECO), grant FPA2010-17915, and by the Junta de Andaluc\'ia, grants FQM 101 and FQM 6552.
\appendix
\section{Induced potential}
\label{AppPot}
Integrating out of the scalar particles induces potential terms of the Higgs doublets.
For the $\textbf{1}_0$ scalar, quadratic terms are also generated by integrating out.
The terms for a Higgs pair, $H_i$ and $H_j$, are as follows:
\begin{align}
 \mcl{L}_M^{(i,j)} = m_0^2 (\la_0^{ii} I_{ii}^0  +\la_0^{ij} I_{ij}^0 +\la_0^{ij\ast} I_{ji}^0 ).
\end{align}
These terms naively generate the Higgs boson whose masses are $O(m_0)$.

The induced quartic couplings can be written as five types of the index combinations introduced in Sect.~\ref{SecInt}.
In the following expressions, we use the notation of the quartic terms such as $H_{ijkl} := (H_i^\dag H_j)(H_k^\dag H_l)$.
They are shown as follows:
\begin{align}
 \mcl{L}_\text{I}^P =&
   ( \la_0^{ii}{}^2 +\la_L^{ii}{}^2 +\la_T^{ii}\la_T^{ii\ast} ) H_{iiii} ,\\
 \mcl{L}_\text{II}^P =&
   (\la_0^{ii} \la_0^{ij} +\la_L^{ii}\la_L^{ij} +\la_T^{ii\ast}\la_T^{ij})H_{iiij} +\hc ,\\
 \mcl{L}_\text{III}^P =&
    \left( (\la_0^{ij}{}^2 +\la_L^{ij}{}^2 +\la_T^{ii\ast} \la_T^{jj}) H_{ijij} +\hc \right) \n &
 +(\la_0^{ii}\la_0^{jj} +\la_S^{ij}\la_S^{ij\ast} -\la_L^{ii}\la_L^{jj} +2\la_L^{ij}\la_L^{ij\ast} +\la_T^{ij}\la_T^{ij\ast}) H_{iijj} \n &
 +(\la_0^{ij}\la_0^{ij\ast} -\la_S^{ij}\la_S^{ij\ast} -\la_L^{ij}\la_L^{ij\ast} +2\la_L^{ii}\la_L^{jj} +\la_T^{ij}\la_T^{ij\ast}) H_{ijji} ,\\
 \mcl{L}_\text{IV}^P =&
  ( \la_0^{ii}\la_0^{jk} +\la_S^{ij\ast}\la_S^{ik} -\la_L^{ii}\la_L^{jk} +2\la_L^{ij\ast}\la_L^{ik} +\la_T^{ij\ast}\la_T^{ik\ast} ) H_{iijk} \n &
 +( \la_0^{ij}\la_0^{ik} +\la_L^{ij}\la_L^{ik} +\la_T^{ii\ast}\la_T^{jk\ast} ) H_{ijik} \n &
 +( \la_0^{ij}\la_0^{ik\ast} -\la_S^{ij}\la_S^{ik\ast} -\la_L^{ij}\la_L^{ik\ast} +2\la_L^{ii}\la_L^{jk\ast} +\la_T^{ij}\la_T^{ik} ) H_{ijki} +\hc ,\\
 \mcl{L}_\text{V}^P =&
  ( \la_0^{ij}\la_0^{kl} +\la_S^{ik\ast}\la_S^{jl} -\la_L^{ij}\la_L^{kl} +2\la_L^{il}\la_L^{jk\ast} +\la_T^{ik\ast}\la_T^{jl} ) H_{ijkl} \n &
 +( \la_0^{ij}\la_0^{kl\ast} +\la_S^{il\ast}\la_S^{jk} -\la_L^{ij}\la_L^{kl\ast} +2\la_L^{ik}\la_L^{jl\ast} +\la_T^{il\ast}\la_T^{jk} ) H_{ijlk} \n &
 +( \la_0^{ik}\la_0^{jl} +\la_S^{ij\ast}\la_S^{kl} -\la_L^{ik}\la_L^{jl} +2\la_L^{il}\la_L^{jk} +\la_T^{ij\ast}\la_T^{kl} ) H_{ikjl} \n &
 +( \la_0^{ik}\la_0^{jl\ast} -\la_S^{il\ast}\la_S^{jk} -\la_L^{ik}\la_L^{jl\ast} +2\la_L^{ij}\la_L^{kl\ast} +\la_T^{il\ast}\la_T^{jk} ) H_{iklj} \n &
 +( \la_0^{il}\la_0^{jk} -\la_S^{ij\ast}\la_S^{kl} -\la_L^{il}\la_L^{jk} +2\la_L^{ik}\la_L^{jl} +\la_T^{ij\ast}\la_T^{kl} ) H_{iljk} \n &
 +( \la_0^{il}\la_0^{jk\ast} -\la_S^{il\ast}\la_S^{jk} -\la_L^{il}\la_L^{jk\ast} +2\la_L^{ij}\la_L^{kl} +\la_T^{ik\ast}\la_T^{jl} ) H_{ilkj} +\hc .
\end{align}
\section{Conditions of the custodial symmetry}
\label{AppCust}
We show the conditions of the custodial symmetry for the Higgs derivative interactions with the couplings of the perturbative UV completions.
The conditions with the Wilson coefficients are given by Ref.~\cite{RefKy}.

For the type I and the type II interactions, the conditions are obviously as written in Sect.~\ref{SecInt}.
The coefficients of $O^T$ operators have to disappear, and all coefficients have to be real.
The conditions given in Sect.~\ref{SecInt} can be obtained with the linear combinations of the following relations after neglecting the imaginary parts of the couplings. 
In terms of the Wilson coefficients, the conditions for the type III effective operators are
\begin{align}
 0 =&  c^T_{iijj} +c^T_{ijji} +c^T_{ijij}, \\
 0 =& 3c^T_{iijj} +c^H_{ijji} -c^H_{ijij},
\end{align}
where $c^H_{\cdots}$ and $c^T_{\cdots}$ are, respectively, coefficients of $O^H_{\cdots}$ and $O^T_{\cdots}$ operators.
Then we obtain the following two relations:
\begin{align}
0=&
  2\frac{2\la_L^{ii} \la_L^{jj} + (\la_L^{ij} +\la_L^{ij\ast})^2}{m_L^2}
 -\frac{4\la_T^{ij} \la_T^{ij\ast} +\la_T^{ii}\la_T^{jj\ast} +\la_T^{jj}\la_T^{ii\ast}}{m_T^2} \n &
 +4\frac{g_0^{ii} g_0^{jj}}{m_0^2}
 -\frac{g_S^{ii} g_S^{jj\ast} +g_S^{jj} g_S^{ii\ast}}{m_S^2},\\
0=&
 -2\frac{(\la_0^{ij} -\la_0^{ij\ast})^2}{m_0^2}
 -4\frac{\la_S^{ij} \la_S^{ij\ast}}{m_S^2}
 +4\frac{(\la_L^{ij} +\la_L^{ij\ast})^2 -\la_L^{ii} \la_L^{jj}}{m_L^2}
 +\frac{-8\la_T^{ij} \la_T^{ij\ast} +\la_T^{ii}\la_T^{jj\ast} +\la_T^{jj}\la_T^{ii\ast}}{m_T^2} \n &
 +12\frac{g_0^{ii} g_0^{jj}}{m_0^2}
 -3\frac{g_S^{ii} g_S^{jj\ast} +g_S^{jj} g_S^{ii\ast}}{m_S^2}.
\end{align}

The type IV operators also require the following two relations:
\begin{align}
 0 =&  c^T_{iijk} +c^T_{ijik} +c^T_{ijki}, \\
 0 =& 3c^T_{iijk} -c^H_{ijik} +c^H_{ijki}.
\end{align}
They can be expressed with the couplings of the perturbative UV completions discussed in this paper as follows:
\begin{align}
0=&
   2\frac{\la_L^{ii} (\la_L^{jk} +\la_L^{jk\ast})
  +(\la_L^{ij} +\la_L^{ij \ast})(\la_L^{ik} +\la_L^{ik\ast})}{m_L^2}
	-\frac{\la_T^{ii} \la_T^{jk\ast} +\la_T^{jk} \la_T^{ii\ast}
  +2(\la_T^{ij} \la_T^{ik\ast} +\la_T^{ik} \la_T^{ij\ast})}{m_T^2},\\
0=&
   -2\frac{(\la_0^{ij} -\la_0^{ij\ast})(\la_0^{ik} -\la_0^{ik\ast})}{m_0^2}
   -2\frac{\la_S^{ij} \la_S^{ik\ast} +\la_S^{ik} \la_S^{ij\ast}}{m_S^2} \n &
	 +\frac{-\la_L^{ii} (\la_L^{jk} +\la_L^{jk\ast})
   +2(\la_L^{ij} +\la_L^{ij\ast})(\la_L^{ik} +\la_L^{ik\ast})}{m_L^2}
   +\frac{-4(\la_T^{ij} \la_T^{ik\ast} +\la_T^{ik} \la_T^{ij\ast})
	 + \la_T^{ii} \la_T^{jk\ast} +\la_T^{jk} \la_T^{ii\ast}}{m_T^2}.
\end{align}

Finally, for the type V operators, the following four relations are known as the conditions to impose the custodial symmetry:
\begin{align}
 0 =&   c^T_{ijkl} +c^T_{ijlk}  +c^T_{ikjl} +c^T_{iklj} +c^T_{iljk} +c^T_{ilkj},\\
 0 =& 3(c^T_{ijkl} +c^T_{ijlk}) -c^H_{ikjl} +c^H_{iklj} -c^H_{iljk} +c^H_{ilkj},\\
 0 =& 3(c^T_{ikjl} +c^T_{iklj}) +c^H_{iljk} -c^H_{ilkj} -c^H_{ijkl} +c^H_{ijlk},\\
 0 =& 3(c^T_{iljk} +c^T_{ilkj}) +c^H_{ijkl} -c^H_{ijlk} +c^H_{ikjl} -c^H_{iklj}.
\end{align}
In terms of the perturbative UV completions, the above relations are written as follows:
\begin{align}
0=&
   \frac{
   (\la_L^{il} +\la_L^{il\ast})(\la_L^{jk} +\la_L^{jk\ast})
  +(\la_L^{ik} +\la_L^{ik\ast})(\la_L^{jl} +\la_L^{jl\ast})
  +(\la_L^{ij} +\la_L^{ij\ast})(\la_L^{kl} +\la_L^{kl\ast})}{m_L^2} \n &
  -\frac{
	  \la_T^{ik} \la_T^{jl\ast} +\la_T^{jl} \la_T^{ik\ast} 
	+ \la_T^{il} \la_T^{jk\ast} +\la_T^{jk} \la_T^{il\ast} 
	+ \la_T^{ij} \la_T^{kl\ast} +\la_T^{kl} \la_T^{ij\ast} }{m_T^2},\\
0=&
  -\frac{
	 (\la_0^{ik} -\la_0^{ik\ast})(\la_0^{jl} -\la_0^{jl\ast})
  +(\la_0^{il} -\la_0^{il\ast})(\la_0^{jk} -\la_0^{jk\ast})}{m_0^2}
  -\frac{
	  \la_S^{ik} \la_S^{jl\ast} +\la_S^{jl} \la_S^{ik\ast} 
  + \la_S^{il} \la_S^{jk\ast} +\la_S^{jk} \la_S^{il\ast} }{m_S^2} \n &
	+\frac{
   2(\la_L^{il} +\la_L^{il\ast})(\la_L^{jk} +\la_L^{jk\ast})
  +2(\la_L^{ik} +\la_L^{ik\ast})(\la_L^{jl} +\la_L^{jl\ast})
	- (\la_L^{ij} +\la_L^{ij\ast})(\la_L^{kl} +\la_L^{kl\ast})}{m_L^2} \n &
	+\frac{
	-2(\la_T^{ik} \la_T^{jl\ast} +\la_T^{jl} \la_T^{ik\ast})
	-2(\la_T^{il} \la_T^{jk\ast} +\la_T^{jk} \la_T^{il\ast})
	+  \la_T^{ij} \la_T^{kl\ast} +\la_T^{kl} \la_T^{ij\ast} }{m_T^2},\\
0=&
   \frac{
  - (\la_0^{ij} -\la_0^{ij\ast})(\la_0^{kl} -\la_0^{kl\ast})
  + (\la_0^{il} -\la_0^{il\ast})(\la_0^{jk} -\la_0^{jk\ast})}{m_0^2}
	+\frac{
  -  \la_S^{ij} \la_S^{kl\ast} -\la_S^{kl} \la_S^{ij\ast} 
	+  \la_S^{il} \la_S^{jk\ast} +\la_S^{jk} \la_S^{il\ast} }{m_S^2} \n &
  +\frac{
	 2(\la_L^{ij} +\la_L^{ij\ast})(\la_L^{kl} +\la_L^{kl\ast})
	+2(\la_L^{il} +\la_L^{il\ast})(\la_L^{jk} +\la_L^{jk\ast})
	- (\la_L^{ik} +\la_L^{ik\ast})(\la_L^{jl} +\la_L^{jl\ast})}{m_L^2} \n &
	+\frac{
	-2(\la_T^{ij} \la_T^{kl\ast} +\la_T^{kl} \la_T^{ij\ast})
  -2(\la_T^{il} \la_T^{jk\ast} +\la_T^{jk} \la_T^{il\ast})
	+  \la_T^{ik} \la_T^{jl\ast} +\la_T^{jl} \la_T^{ik\ast} }{m_T^2},\\
0=&
   \frac{
    (\la_0^{ij} -\la_0^{ij\ast})(\la_0^{kl} -\la_0^{kl\ast})
  + (\la_0^{ik} -\la_0^{ik\ast})(\la_0^{jl} -\la_0^{jl\ast})}{m_0^2}
	+\frac{
	 \la_S^{ij} \la_S^{kl\ast} +\la_S^{kl} \la_S^{ij\ast}
	+\la_S^{ik} \la_S^{jl\ast} +\la_S^{jl} \la_S^{ik\ast}}{m_S^2} \n &
	+\frac{
	 2(\la_L^{ij} +\la_L^{ij\ast})(\la_L^{kl} +\la_L^{kl\ast})
	+2(\la_L^{ik} +\la_L^{ik\ast})(\la_L^{jl} +\la_L^{jl\ast})
	- (\la_L^{il} +\la_L^{il\ast})(\la_L^{jk} +\la_L^{jk\ast})}{m_L^2} \n &
	+\frac{
  -2(\la_T^{ij} \la_T^{kl\ast} +\la_T^{kl} \la_T^{ij\ast})
  -2(\la_T^{ik} \la_T^{jl\ast} +\la_T^{jl} \la_T^{ik\ast})
	+  \la_T^{il} \la_T^{jk\ast} +\la_T^{jk} \la_T^{il\ast} }{m_T^2}.
\end{align}
The real parts of $\la_0$ and $g_L$ do not appear in any relations, so that they always respect the custodial symmetry.
\end{document}